# Examining Data Imbalance in Crowdsourced Reports for Improving Flash Flood Situational Awareness

*Miguel Esparza[1*], Hamed Farahmand[1], Samuel Brody[2], Ali Mostafavi[1]*

**Affiliations:**
[1]M.Sc. Student, Zachry Department of Civil and Environmental Engineering, Urban Resilience. AI Lab, Texas A&M University, College Station; email: mte1224@tamu.edu
*Corresponding Author
[2]Ph.D. Candidate, Zachry Department of Civil and Environmental Engineering, Urban Resilience. AI Lab, Texas A&M University, College Station; email: hamedfarahmand@tamu.edu
[3]Associate Professor, [2]Institue for a Disaster Resilient Texas, Department of Marine Sciences, Texas A&M University at Galveston, Galveston; email: brodys@tamug.edu
[4]Associate Professor Urban Reslience.AI Lab, Zachry Department of Civil and Environmental Engineering, Texas A&M University, College Station; email: amostafavi@civil.tamu.edu

## ABSTRACT

The use of crowdsourced data has been finding practical use for enhancing situational awareness during disasters. While recent studies have shown promising results regarding the potential of crowdsourced data (such as user-generated flood reports) for flash flood mapping and situational awareness, little attention has been paid to data imbalance issues that could introduce biases in data and assessment. To address this gap, in this study, we examine biases present in crowdsourced reports to identify data imbalance with a goal of improving disaster situational awareness. Three biases are examined: sample bias, spatial bias, and demographic bias. To examine these biases, we analyzed reported flooding from 3-1-1 reports (which is a citizen hotline allowing the community to report problems such as flooding) and Waze reports (which is a GPS navigation app that allows drivers to report flooded roads) with respect to FEMA damage data collected in the aftermaths of Tropical Storm Imelda in Harris County, Texas, in 2019 and Hurricane Ida in New York City in 2021. First, sample bias is assessed by expanding the flood-related categories in 3-1-1 reports. Integrating other flooding related topics into the Global Moran's I and Local Indicator of Spatial Association (LISA) revealed more communities that were impacted by floods. To examine spatial bias, we perform the LISA and BI-LISA tests on the data sets—FEMA damage, 3-1-1 reports, and Waze reports—at the census tract level and census block group level. By looking at two geographical aggregations, we found that the larger spatial aggregations, census tracts, show less data imbalance in the results. Finally, one-way Analysis of variance (ANOVA) test performed on the clusters generated from the BI-LISA shows that data imbalance exists in areas where minority populations reside. Through a regression analysis, we found that 3-1-1 reports and Waze reports have data imbalance limitations in areas where minority populations reside. The findings of this study advance understanding of data imbalance and biases in crowdsourced datasets that are growingly used for disaster situational awareness. Through addressing data imbalance issues, researchers and practitioners can proactively mitigate biases in crowdsourced data and prevent biased and inequitable decisions and actions.

# 1. INTRODUCTION

## 1.1 Background

Natural hazards are increasing in frequency and intensity, causing loss of life, inflicting damage to the built environment, and imposing severe economic repercussions. Flooding in particular, has caused extensive impacts in urban areas over the past decade (Aerts et al. 2018; Dong et al. 2019; Jonkman 2005). Enhanced situational awareness is critical to responding to natural hazards, such as flooding events. One way to enhance situational awareness is to improve the extent of information used in disaster management. Emerging data sensing technologies are attracting increasing attention to effectively map the status of hazard impacts and their spatial and temporal characteristics. Specifically, using crowdsourced data to improve situational awareness in disasters, such as floods, has been growing in recent years (Paradkar et al. 2022).

The nearly ubiquitous use of digital devices and crowdsourced data platforms offers a wealth of data whose analysis enhances sensing of disaster situations. Crowdsourced data (such as user-generated data from 3-1-1 and crowdsourced traffic updates from Waze reports) enrich information capacity and rapid interactivity by allowing the general population to share critical information regarding the disaster, thus enhancing situational awareness (Ahmouda, Hochmair, and Cvetojevic 2019; Cvetojevic and Hochmair 2018; Sutton et al. 2015; Yang et al. 2019). Moreover, by interpreting the spatial patterns of crowdsourced data during a disaster, emergency managers and city officials can identify the extent of impacts and prioritize resource allocation and response activities. While the use of crowdsourced data has been growing in recent years, limitations, such as potential data imbalances still need to be addressed. Certain communities or geographical locations receive more attention on crowdsourced platforms, which causes bias in these datasets. The objective of this research is to examine data bias and potential data imbalances in disaster-related crowdsourced data to enhance situational awareness for improving disaster response.

## 1.2 Types of Bias in Crowdsourced Data

Crowdsourced and social media data has been under criticisms due to the disparities that the information from these data sets (Fan et al. 2020; Srivastava and Mostafavi 2018). For example, crowdsourced data tends to represent privileged rather than underprivileged communities—which are more strongly impacted by a disaster—due to the social demographics of these two groups on crowdsourced platforms being vastly different (Dargin, Fan, and Mostafavi 2021; Zhang et al. 2019). Such a disparity is a result of data bias. To mitigate data bias (in domains other than disasters), researchers have extensively studied how data aggregation enhances the results provided by the data. However, limited attention has been paid to data imbalance and biases in datasets (specifically crowdsourced data) vis-à-vis disasters. To address this knowledge gap, we examine types of bias that occur in crowdsourced data and how to mitigate these biases for improving situational awareness. The following sections give an overview on the types of bias that the research aims to assess: *section 1.2.1* examines sample bias in crowdsourced data, *section 1.2.2* discusses spatial bias and *section 1.2.3* covers demographic bias.

### *1.2.1 Sample bias in crowdsourced data*

Sample bias, which occurs when collecting crowdsourced data, can skew the information conveyed by the data. The majority of studies focus on sample bias in Twitter data. For example, geotagged tweets suffer from a form of sample bias as these tweets form only a small sample of Twitter data that may hide valuable information. Only about 1.0% to 1.5% of tweets are actually geotagged (Ave et al. 2013), and fewer than 0.42% of all tweets are associated with accurate geospatial information (Cheng, Caverlee, and Lee 2010). The lack of representation of geotagged tweets compared with total tweets highlights the severity of sample bias; therefore, studies have addressed sample bias when collecting datasets, such as disaster-related tweets.

For example, Morstatter and Lui (2013) examined how to mitigate data bias, particularly sample bias, within Twitter's API system. Collecting Twitter data from their API presents a challenge for researchers as it limits the sample to only 1% of all Twitter data. To address this bias, the researchers developed several algorithms that take advantage of Twitter hashtags, such as a spectral clustering algorithm that incorporates similarities between Tweets, to work around the 1% sample limit. These algorithms were tested with Twitter's sample API. These algorithms were found to be effective in mitigating sample bias (Morstatter et al. 2013). Another form of sample bias that can emerge from collecting Twitter data is mapping tweets without examining their context; crucial information regarding preparedness and damage reports is lost from the tweets. Acknowledging sample bias, Wang et al. (2019) proposed to use machine learning to examine the context of tweets to properly gather this information and identify which areas' residents needed rescue. The context of tweets that their machine learning model captured after Hurricane Sandy holistically allowed the researchers to identify the severely impacted areas from the larger sample of tweets. This larger sample allowed them to find that socially vulnerable communities did not receive the resources or attention needed to properly recover from the disaster, as such information could not be captured without looking at the context and a larger sample of tweets (Wang et al. 2019). Samuels et al. (2020) assessed that social media traffic dropped off during Hurricane Sandy to address sample bias in Twitter data. Their analysis reported correlations between infrastructure damage and the magnitude of social media activity changes from a defined steady state, which allowed them to monitor social media drop-offs. The analysis showed the existence of sample bias introduced by the transition of a normal state to a crisis state during a disaster. Moreover, failing to recognize social media drop-offs exacerbates sample bias and puts vulnerable populations at risk (Samuels, Taylor, and Mohammadi 2020). Despite the growing recognition of the importance of understanding sample bias in crowdsourced data in disasters, most of existing studies focus on twitter data, while limited attention is paid to other crowdsourced data that are commonly used for disaster situational awareness.

### 1.2.2 Spatial bias in crowdsourced data

Spatial bias can limit the knowledge that can be acquired when examining spatial patterns of crowdsourced data. This bias can occur when assuming spatial concentrations of crowdsourced data need relief after a disaster because it over-simplifies the way these data can be used to help vulnerable populations recover.

The geographical scale that is used for aggregation is one factor that influences of spatial bias in crowdsource data. For example, the uneven spatial distributions of census tracts and ZIP Codes have been examined for data bias (Grubesic 2008; Grubesic and Matisziw 2006; Jelinski and Wu 1996; Saib et al. 2014). To address this spatial bias, Samuels et al. (2022) divided the Harris County region into even hexagonal shapes, rather than using census tracts, to examine how a range of different geographical scales impacts the correlation between Twitter and Hurricane Harvey damage data. The power law relationship and strength of correlation between Twitter and Hurricane Harvey damage data indicated that larger spatial scales provide improved results from disaster-related content on social media (Samuels et al. 2022). Additionally, Graham et al. (2014) identified a discrepancy in correlation values between Twitter activity and Hurricane damage at varying geographical scales, noting the apparent necessity of including scale as a factor in any correlation analysis. Moreover, their spatial analysis revealed little Twitter activity in Staten Island during Hurricane Sandy, despite half the deaths related to Hurricane Sandy came from Staten Island. The fact that social media data did not identify an area that faced extreme devastation from Hurricane Sandy highlights the negative impact that spatial bias has when using crowdsourced data for disaster assessments (Graham et al. 2014). Geotagged tweets have an inherited spatial bias as they may not accurately represent the spatial location of a disaster, yet a number of studies have used geotagged tweets to examine the impacts of flooding events (Li et al. 2018; Martín et al. 2020; Yin et al. 2015). To mitigate the spatial bias of geotagged tweets, Fan et al. (2020) used a content-based analysis that focused on the context of the tweets rather than merely treating tweets as a spatial point. Through the content-based analysis, the research assessed the relationship between population size of city centers and number of different types of Tweets with a power law relationship. The power law relationship found in the study indicates that as the number of people grow within a city, the social media usage grows as well, which gives preferential treatment in the spatial clusters of people who can use social media during a disaster (Fan et al. 2020).

### *1.2.3 Demographic bias in crowdsourced data*

Demographic bias refers to the lack of representation of socially vulnerable groups in crowdsourced data. Real-time information from users impacted by a disaster derived from crowdsourced data improves disaster situational awareness. However, access to digital devices is not uniform, instilling biases in crowdsourced data that may exacerbate the recovery process for communities with limited access. Sociodemographic factors that may underlie bias include: gender inequalities in reporting patterns, as women are typically not accurately represented on crowdsourced platforms; social capital reporting patterns, as lower-income groups often do not have access to methods of reporting, such as a smart phone (Ellison, Steinfield, and Lampe 2007; Jung, Ozkaya, and Larose 2014; Warren, Sulaiman, and Jaafar 2015); racial inequalities, which prevent minorities from being represented on crowdsourced platform (Barberá 2016; Cutter, Boruff, and Shirley 2003; Murthy, Gross, and Pensavalle 2016), and age reporting, as elderly populations may not be familiar with ways to report or the younger populations not having access to a way to report (Brandtzæg 2012; Pfeil, Arjan, and Zaphiris 2009). Thus, social demographic traits have an impact on representation on crowdsourced platforms; therefore, demographic bias needs to be examined in crowdsourced data to enable data-based equitable decisions and actions.

Several studies of demographic bias in crowdsourced data and have examined the extent to which socially vulnerable groups suffer from unequal representation during a disaster. For example, during Typhoon Haiyan, in the Philippines, Mandianou (2015) examined how digital inequalities on social media hindered the recovery of low-income participants. A year after the typhoon, the researchers found that many low-income participants were living in temporary housing without stable forms of income, while middle-class social media users recovered well and in some cases strengthened their status due to their voices being heard on social media (Madianou 2015). Forati and Ghose (2022) examined social and spatiotemporal inequalities in the use of Twitter after Hurricane Irma in the Pinellas County, Florida. Their geographic weighted regression model assessed the relationship between racial minorities in impoverished neighborhoods and social media content related to the disaster. They found that poorer inland communities were absent on social media (Forati and Ghose 2022). Yu Xiao et al. (2015) examined the spatial heterogeneity in the generation of tweets after Hurricane Sandy to address which social groups were omitted from social media content during a disaster. They found that a community's socioeconomic factors are more important than population size and damage levels in predicting disaster-related tweets (Xiao, Huang, and Wu 2015).

## 2 RESEARCH FRAMEWORK AND HYPOTHESIS

To assess data imbalance issues in crowdsourced data for improving situational awareness during disasters, this study focuses on three research questions addressed using the framework illustrated in Fig. 1. We examine two primary crowdsourced datasets: 3-1-1 reports and Waze reports. These two crowdsourced datasets are used for situational awareness of extreme weather and flash flood events.

1. **Sample bias:** To what extent does expanding crowdsourced data to include more flooding-related impacts reduce sample bias in 3-1-1 reports?

To address research question 1, the Global Moran's I and Local Indicator of Spatial Association (LISA) test were used to compare the spatial patterns of different topic categories of 3-1-1 reports. The rationale is that holistically looking at spatial patterns of 3-1-1 reports by expanding the categories (beyond street flooding) could reveal areas that experienced extensive flooding induced by the disasters being studied.

2. **Spatial bias:** To what extent does the spatial level of aggregation for 3-1-1 and Waze report affect the insights obtained from crowdsourced data for flood damage assessment?

The LISA and BI-Local Indicator of Spatial Association (BI-LISA) tests are performed at two geographical aggregations, census tracts (the larger aggregation) and census block groups (the smaller aggregation) to examine the impacts of spatial bias. By comparing LISA maps of damaged areas and crowdsourced data, we identified spatial patterns at larger geographical aggregations and compared those patterns at a smaller aggregation to assess the extent to which information was lost. Once general patterns are obtained through the LISA test, we performed the BI-LISA test to identify areas with data imbalance clusters (areas with low damage, yet high reporting or vice versa) and compare these imbalance clusters at both geographical aggregations. Next, to understand why these data imbalance clusters exists, we performed the one-way ANOVA test on the social demographic traits for each cluster category produced by the BI-LISA test. We hypothesize that minority populations reside in spatial clusters with low FEMA damage, yet high crowdsourced reports or vice versa for both flooding events and both geographical aggregations.

3. **Demographic bias:** What social demographic traits explain the variation in the number of crowdsourced reports and how do these characteristics affect crowdsourced reporting imbalances?

To address research question 3, linear regression models were developed to assess the extent to which social demographic traits impacted crowdsourced reporting patterns. The independent variables for this analysis are social demographic traits and FEMA damage claims, while the dependent variable are the crowdsourced reports. Moreover, the data used in the regression models were aggregated at the census tract level and census block group level. The research hypothesizes that social demographic traits, such as minority status, are correlated with reporting patterns more than population size.

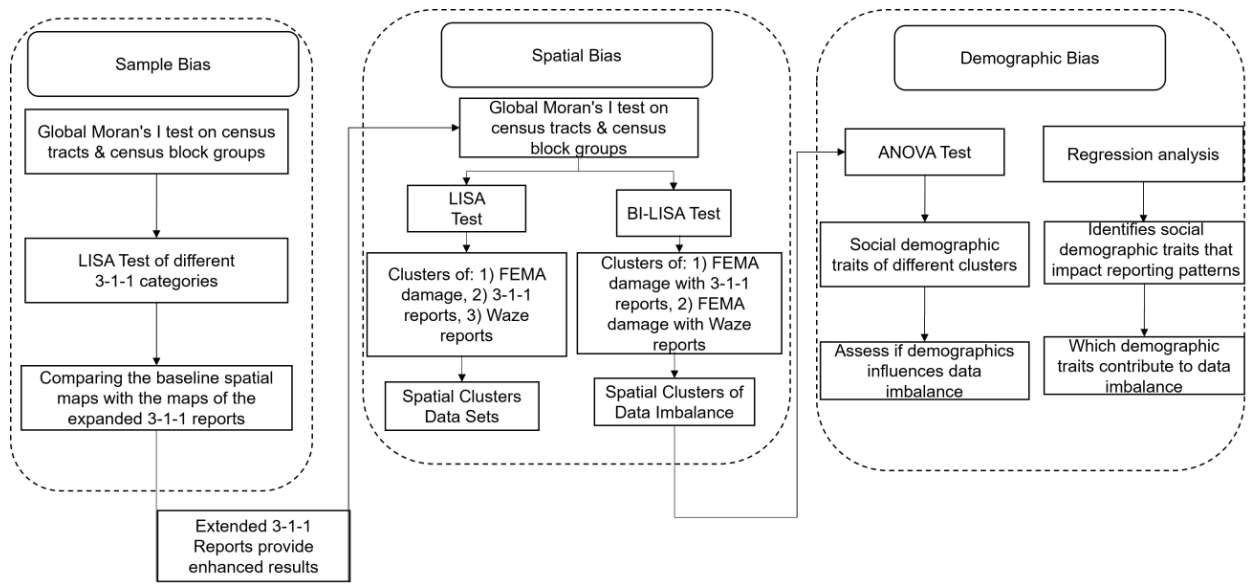

**Fig. 1. Analytical framework for assessing data imbalance in crowdsourced data for improving disaster situational awareness**

## 3  STUDY AREA AND DATA DESCRIPTION

In this study, we used 3-1-1 and Waze reports related to major flash flood events in the United States: Tropical Storm Imelda in Houston and Hurricane Ida in New York City. Tropical Storm Imelda occurred on September 19, 2019 and lasted through September 21, 2019. During the flooding event, more than 50 inches of rain fell in the Houston and southeast Texas area. The disaster was the fifth-wettest tropical cyclone in the United States. Hurricane Ida made landfall on August 29, 2021 and lasted through September 9, 2021. The Category 4 hurricane with 150 mph windspeed devastated almost one million homes and business along the US Gulf Coast. When it reached New York City, Ida caused severe flash flooding in multiple parts if the city.

We collected 3-1-1 and Waze reports for these two flood events for the respective study areas. 3-1-1 is a citizen hotline that allows the community to report problems, such as flooding, around their community. Waze is a GPS navigation app that allows drivers to report road conditions, including flooding. Both crowdsourced datasets are characterized by a location and timestamp. These crowdsourced data are shown to be reliable for flash flood mapping (Yuan et al. 2021, 2022). For Tropical Storm Imelda, the baseline category was Flooding, and the expanded categories were drainage, storm debris collection, and crisis cleanup. For Hurricane Ida the baseline category was storm, and the expanded category was sewer. These 3-1-1 reports were normalized based on housing units, while Waze reports were normalized based on road lengths. The FEMA damage and crowdsourced datasets were gathered from September 19, 2019, to September 21, 2019, for Tropical Storm Imelda, and August 29, 2021, to September 9, 2021, for Hurricane Ida. We also gathered 2018 census data from the American Community Survey at the census tract and census block group level for the following social parameters: housing unit population, minority, single-parent households, no education (NOEDU), and poverty status. To account for the impact of flooding imposed by Tropical Storm Imelda and Hurricane Ida, the research used FEMA damage claims (FEMA 2019, 2021). The FEMA damage data was gathered on January 28, 2022. The FEMA damage data is aggregated at the census tract level and to uphold FEMA privacy guidelines, the only geographical information given were the census tracts in which that the claims were reported. Therefore, to gather the FEMA damage data at the census block group level, the research multiplied the number of claims at the census tract level by a weighted average ratio of census tract area divided by each census block group area (United States Census Bureru 2018). All datasets were normalized by the following equation (1), where $q$ is the normalized parameter. $x_i$ is the individual data point, and the $minimum\ (x)$ is the minimum value of the whole dataset, while the $maxmium(x)$ is the maximum value of the whole dataset.

$$q = \frac{x_i - minimum\ (x)}{maxmium(x) - minium(x)} \qquad (1)$$

## 4  METHODOLOGY

To address sample bias, we performed the Global Moran's I and LISA test on a baseline topic of 3-1-1 reports, flood and storm, and on expanded categories that associate with debris clean up, sewer, and drainage. The LISA test revealed areas that experienced extensive flooding by expanding the topics that are studied. After sample bias is addressed, we extended the LISA results to FEMA damage claims and Waze reports in addition to the expanded 3-1-1 reports to address spatial bias. The research performed the LISA and BI-LISA test at the census tract and census block group level to assess how geographical aggregation impacts the reliability of damage assessments using crowdsourced data. Next, to assess what social demographic groups are impacted by data imbalance, the one-way ANOVA test was performed on the four types of cluster categories generated through the BI-LISA statistic. To further assess demographic bias, regression analysis was performed to gain a better understanding of how social demographic traits can explain the variation in the number of crowdsourced reports in flood impacted areas.

## 4.1 Spatial Analysis for Addressing Sample Bias and Identifying Geographical Disparities

Spatial analysis is performed on the FEMA damage, census data, 3-1-1 reports, and Waze reports to examine sample bias and spatial bias. The objective of spatial analysis is to: 1) examine the extent to which expanding 3-1-1 topic categories enhance flash flood mapping results; and 2) assess how spatial aggregation impacts the analysis results of flash flood mapping with crowdsourced data.

The first step of spatial analysis is to perform the Global Moran's I test on the two crowdsourced datasets, 3-1-1 reports and Waze reports, and FEMA damage claims. The Global Moran's I statistic is computed to determine if neighboring census tracts or census block groups share a similar status with crowdsourced reporting and if the damaged areas are near each other. The null hypothesis for the Global Moran's I tests holds that the attributes being analyzed are randomly distributed among the features in the study area. Therefore, a p-value of less than 0.05 indicates the emerging spatial patters are not random. The Global Moran's I statistic can be computed with the equation 2 (Moran 1948).

$$I = \frac{n}{S_0} \frac{\sum_{i=1}^{n} \sum_{j=1}^{n} w_{ij} \cdot z_i \cdot z_j}{\sum_{i=1}^{n} z_i^2} \tag{2}$$

In this equation $z_i = x_i - \bar{x}$, where $\bar{x}$ is the mean of variable $x$. The weight matrix, $w_{ij}$, is a spatial matrix of binary code to indicate adjacent neighbors. The number of features is represented by $n$ and $S_0 = \sum_{i=1}^{n} \sum_{j=1}^{n} w_{i,j}$, which is the summation of all spatial weights.

In the next step, the study area is decomposed into local clusters due to the statistical significance of the Global Moran's I test. To break down the global space, the LISA statistic is used for identifying clusters which have a similar number of reporting or damage patterns. Obtaining different spatial clusters through spatial analysis can help identify data imbalance by examining the spatial patterns and social traits of the statistically significant clusters. The LISA statistic can be calculated by equation 3 (Anselin 1995).

$$I_i = \frac{z_i}{\frac{\sum_i^n (z_i)^2}{n}} \cdot \sum_j^n w_{ij} \cdot z_j \tag{3}$$

This equation is similar to equation 2, $z_i = x_i - \bar{x}$, where $\bar{x}$ is the mean of variable $x$, $w_{ij}$ is the spatial weight matrix, and $n$ is the number of observations. Essentially, equation 3 standardized the value x for observation, $i$, and the neighboring observations, $j$, to determine if $i$ and $j$ are high or low relative to the mean. This measure allows for the decomposition of the Global Moran's I index into individual observations to assess the significance and contribution of the local clusters. The results of this test produced statistically significant polygons that formed clusters of disruption. The cluster categories are as follows:

- High–high: Areas that have a high means and are surrounded by other geographical areas with similar values.
- Low–low: Areas that have a low means and are surrounded by other geographical areas with similar values
- High–low: Areas that have a high means and are surrounded by other geographical areas with lower values
- Low–high: Areas that have a low means and are surrounded by other geographical areas with higher values

Another component of the spatial analysis is using a multivariate analysis to assess data imbalance through the BI-LISA test. The BI-LISA test measures the degree which the value for a given variable at a location is correlated with its neighbors for a different value. Moreover, it correlates variable $x_i$, variable 1, with the spatial lag of another variable $\sum_j^n w_{ij} y_j$, where $w_{ij}$ is the spatial weight matrix and $y_j$ is the second variable being studied. This application allows the research to examine the imbalances between damage data ($x_i$,) and the spatial lag of crowdsourced reporting ($\sum_j^n w_{ij} y_j$). For example, if a cluster has low damage, yet high reporting then this cluster experienced a form of data imbalance. The equation of the BI-LISA test is shown in equation 4.

$$I_i^B = \frac{\sum_i(\sum_j^n w_{ij} y_j \cdot x_{i)}}{\sum_i(x_i^2)} \tag{4}$$

## 4.2 One-way ANOVA Test for Assessing Social Traits of Spatial Clusters

To examine the underlying characteristics of the spatial clusters, we performed the one-way ANOVA test. The ANOVA model is commonly used in research for comparing multiple groups. In this case, the ANOVA test is performed on the social demographic characteristics of the four categories that are generated by the two BI-LISA tests: 1) FEMA damage with 3-1-1 reports and 2) FEMA damage with Waze reports. This analysis reveals data imbalances (i.e., high damage but low reporting and vice versa) and assesses why these data imbalance clusters are generated by looking at the statistically significant social demographic traits and the distribution of certain social groups that reside in data imbalance clusters. The equation for the ANOVA test is shown below (Tian et al. 2018).

$$Y_{ij} = \mu_i + \varepsilon_{ij}, \varepsilon_{ij} \sim N(0, \sigma^2), 1 < j < n_i, 1 < i < I \tag{5}$$

The equation considers a continuous outcome of $Y$, and $I$ denotes the number of groups being examined. The main interest is comparing the mean of $Y$ across the $I$ groups. Moreover, $Y_{ij}$ is the outcome from the $j$th subject within the $i$th group; $\mu_i = E(Y_{ij})$ is the mean of the $i$th group; $\varepsilon_{ij}$ is the error term, $N(0, \sigma^2)$ denotes the normal distribution with mean $\mu$ and $\sigma^2$; and $n_i$ is the sample size of the $i$th group.

## 4.3 Regression Models for Assessing Social Characteristics Impact on Reporting

Regression models are implemented to examine the extent to which social demographic traits explain the variation of crowdsourced reporting by examining the relationship of crowdsourced data with census data and FEMA damage data. The census data that the research examines are in terms of percentages and are as follows: population, poverty, populations with no education (NOEDU), single-parent households, and minority status. The independent variables are census data and FEMA damage data, while two crowdsourced data sets, 3-1-1 reports, and Waze reports, are the dependent variable. Moreover, all variables are aggregated based on census tracts and census block groups. The following models were implemented to assess the degree that social demographic traits explain the extent of crowdsourced reporting and which demographic traits experience demographic bias:

Model 1: $Crowdsourced\ Data = \alpha + \beta_1 Population + e$ (6)

Model 2: $Crowdsourced\ Data = \alpha + \beta_1 Population + \beta_2 FEMA\ Claim + e$ (7)

Model 3: $Crowdsourced\ Data = \alpha + \beta_1 Poverty\% + \beta_2 NOEDU\% + \beta_3 Single\ Parent\ Household\% + \beta_4 Minority\% + e$ (8)

Model 4: $Crowdsourced\ Data = \alpha + \beta_1 Population + \beta_2 FEMA\ Claim + \beta_3 Poverty\% + \beta_4 NOEDU\% + \beta_5 Single\ Parent\ Household\% + \beta_6 Minority\% + e$ (9)

Model 1 includes only the population parameter as a baseline for examining how crowdsourced reporting is impacted by population size. Model 2 adds the damage parameter to assess how the availability of material of reporting (damaged homes) affects the reporting patterns. Model 3 includes only the social demographic traits as the independent variable. By comparing Model 3 with Models 1 and 2, the research is able to compare the variation of crowdsourced reports between socially vulnerable areas with populous and damaged areas. Model 4 includes all independent variables to holistically assess the variation in crowdsourced reporting. By examining all four models, the research addresses demographic bias by examining the extent that different social traits impact reporting patterns.

# 5 RESULTS

We first present an overview of the descriptive statistics and descriptive maps of the data. Next, we discuss the sample bias results in data by expanding the categories of 3-1-1 reports that are examined. This addresses the first research questions. The next set of results examine spatial bias by looking at the spatial concentration of FEMA damage and crowdsourced reporting using the LISA statistic and BI-LISA statistic. Moreover, to gain a better understanding of which social groups suffered from data imbalances, the one-way ANOVA test is performed to compare the different social demographic groups on the spatial clusters generated by the BI-LISA test. Finally, we present the results related to the demographic attributes that explain crowdsourced reporting variation, and which social demographics suffer from demographic bias.

## 5.1 Descriptive Analysis

Fig. 2 shows the descriptive maps of Harris County (Houston metro area) during Tropical Storm Imelda for FEMA damage, 3-1-1 reports, and Waze reports at the census tract (2(a)–2(b)) and the census block level (2(d)–2(f)). The descriptive maps show that, overall, the distribution of all datasets at both geographical aggregations are consistent. For example, in Fig. 2 shows concentrations of high damage (2(a)), 3-1-1 reports (2b), and Waze reports (2(c)) in the downtown area at the census tract level and census block group level. Fig. 3 presents the descriptive analysis of New York during Hurricane Ida for the same data sets. Notably, there are an abundance of clusters with high Waze reports (3(c)), while FEMA damage (3(a)) and 3-1-1 reports (3(b)) have similar reporting patterns. To ensure the patterns in Fig. 2 and Fig. 3 are consistent at the census tract and census block group level, we performed the LISA and BI-LISA tests, whose results are presented next.

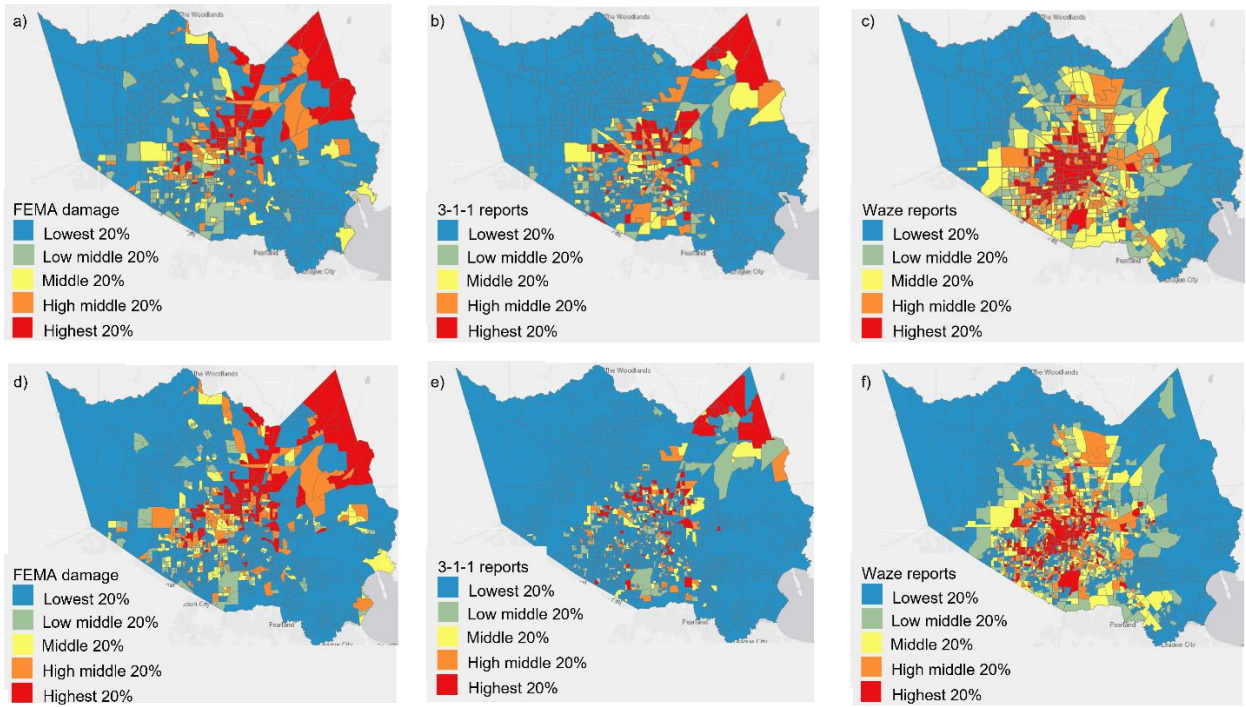

**Fig. 2. Descriptive map of (a and d) FEMA damage, (b and e) 3-1-1 reports, and (c and f) Waze reports at the census tracts (a–c) and census block group (d–f) for Harris County during Tropical Storm Imelda.**

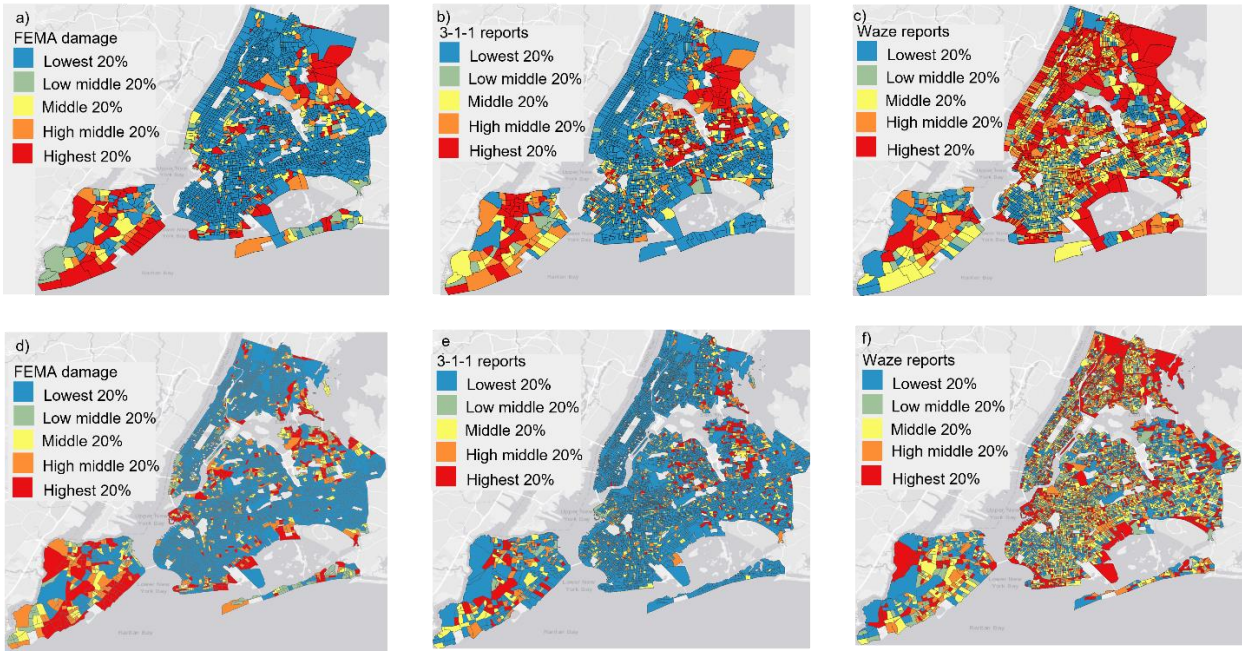

**Fig. 3. Descriptive map of (a and d) FEMA damage, (b and e) 3-1-1 reports, and (c and f) Waze reports at the census tracts (a–c) and census block groups (d–f) for New York during Hurricane Ida.**

Table 1 and Table 2 presents the descriptive statistics of normalized FEMA damage, 3-1-1 reports, Waze reports, and census data at the census tract and census block group level for Harris County and New York, respectively. For Harris County and New York, it is seen that the minority populations have the largest means among the social variables examined. Moreover, at the census block group level, there are smaller mean values for FEMA damage, 3-1-1 reports, and Waze reports when compared to the census tract level.

**Table 1. Tropical Storm Imelda descriptive statistics.**

|  | Census tract | | | | Census block group | | | |
| --- | --- | --- | --- | --- | --- | --- | --- | --- |
|  | Minimum | Maximum | Mean | Standard deviation | Minimum | Maximum | Mean | Standard deviation |
| **FEMA damage** | 0 | 191 | 4.07 | 15.6 | 0 | 191 | 4.38 | 0.17 |
| **3-1-1 report** | 0 | 273 | 1.86 | 12.1 | 0 | 196 | 5.75 | 0.681 |
| **Waze report** | 0 | 116 | 6.6 | 11.2 | 0 | 92 | 2.43 | 5.29 |
| **Population** | 150 | 22567 | 5634 | 3119 | 83 | 22394 | 2145 | 1791 |
| **Poverty** | 0 | 78 | 18.6 | 13.0 | 0 | 5204 | 504 | 445 |
| **NOEDU** | 0 | 66.7 | 21.68 | 16.5 | 0 | 13459 | 1450 | 1135 |
| **Minority** | 0 | 100 | 68.4 | 26.6 | 0 | 1 | 0.358 | 0.244 |
| **Single parent household** | 0 | 45 | 11.8 | 6.9 | 0 | 962 | 101 | 103 |

**Table 2. Hurricane Ida descriptive statistics.**

|  | Census tract | | | | Census block group | | | |
| --- | --- | --- | --- | --- | --- | --- | --- | --- |
|  | Minimum | Maximum | Mean | Standard deviation | Minimum | Maximum | Mean | Standard deviation |
| **FEMA damage** | 0 | 101 | 0.78 | 3.55 | 0 | 101 | 0.9 | 4.19 |
| **3-1-1 report** | 0 | 127 | 5.19 | 8.32 | 0 | 230 | 15.4 | 62 |
| **Waze report** | 0 | 1289 | 43.56 | 98.45 | 0 | 3252 | 15.4 | 62.3 |
| **Population** | 20 | 28272 | 3933 | 2202 | 15 | 8830 | 1349 | 650 |
| **Poverty** | 0 | 5017 | 743 | 713 | 0 | 548 | 47.8 | 58.3 |
| **NOEDU** | 0 | 4318 | 515 | 478 | 0 | 7300 | 893 | 477 |
| **Minority** | 0 | 25738 | 2715 | 2008 | 0 | 6344 | 781 | 596 |
| **Single parent household** | 0 | 1518 | 143 | 150 | 0 | 1222 | 118 | 103 |

## 5.2 Examining Sample Bias

To address sample bias, we examined the extent to which inclusion of other categories of 3-1-1 reports beyond street flooding could change the spatial results. Sample bias was not examined for Waze reports as the data had only one category available, flooded roads. 3-1-1 reports come from a citizen hotline that allows the community to report a wide array of problems in their area and provides different categories of reports. When using 3-1-1 reports to enhance flash flood situational awareness, most studies (Dong et al. 2022) have used flood-related topics. Fig. 4 presents a descriptive bar graph of the different flood-related categories for Tropical Storm Imelda (4(a)) and Hurricane Ida (4(b)). The most intuitive category to examine are Flooding (4(a)) and Storm (4(b)). However, Fig. 4 shows that if analyses only focus on those two topics- Flooding and Storm, then important insight is lost. For example, during Tropical Storm Imelda, only 407 3-1-1 reports were under the category of Flooding for the studied period, whereas Drainage (705), Storm Debris Collection (322), and Crisis Cleanup (181) made up 1,208 of the 1,615 flood related reports (74.8%), respectively. The rationale behind considering these expanded topics for Tropical Storm Imelda is that Drainage considers the overflow of drainage systems when intense flooding occurs, while Storm Debris Collection and Crisis Cleanup consider the debris that impose on the community and need to be cleared. During Hurricane Imelda, Storm is the most initiative category from the New York 3-1-1 hotline; however, Sewer made up 8,889 out of 11,102 flood related reports (80%). The reasoning for selecting Sewer as an expanded category is that New York's sewer system may not be able to handle the intense flooding that Hurricane Ida imposed.

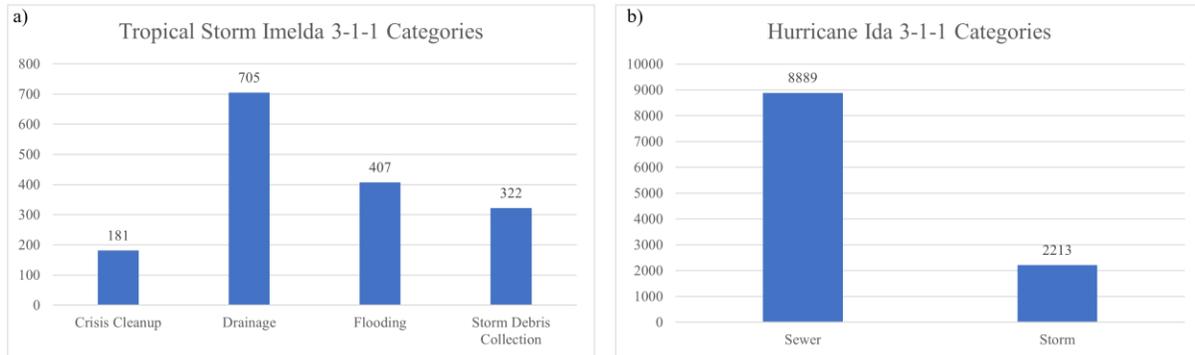

**Fig. 4. Descriptive bar graph of 3-1-1 categories related to (a) Tropical Storm Imelda and (b) Hurricane Ida.**

Table 3 presents the Global Moran's I of only Flooding and Storm, referred to as *Baseline Reports,* and then the Global Moran's I of the all 3-1-1 categories for both flooding events, referred to as *Expanded Reports.* The results show that the *Expanded Reports,* have a stronger Moran's I at both the census tract and census block group level for Tropical Storm Imelda. Hurricane Ida had a stronger Moran's I only at the census block group level. The stronger Moran's I implies that neighboring areas shared a similar experience in terms of 3-1-1 reports during the two flooding events.

**Table 3. Global Moran's I results of intuitive reports and expanded reports.**

| Event | Tropical Storm Imelda | | Hurricane Ida | |
|---|---|---|---|---|
| **Spatial unit** | **Census tract** | **Census block group** | **Census tract** | **Census block group** |
| **Baseline reports** | 0.247*** | 0.229*** | 0.448*** | 0.016** |
| **Expanded reports** | 0.596*** | 0.367*** | 0.034*** | 0.053** |

Fig. 5 compares the spatial patterns of the *Baseline Reports* at the census tract level (5(a)) and at the census block group level (5(c)) with the *Expanded Reports* (5(b)) at the census tract level; and at the census block group level (5(d)). It is seen that the spatial patterns of the *Expanded Reports* provide more reliable insights at both geographical scales as there are more statistically significant clusters due to the research addressing sample bias by expanded the 3-1-1 categories.

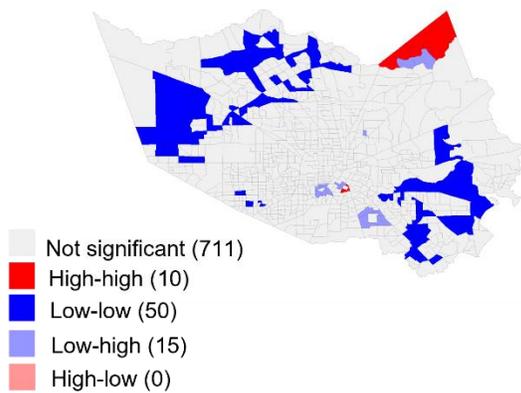
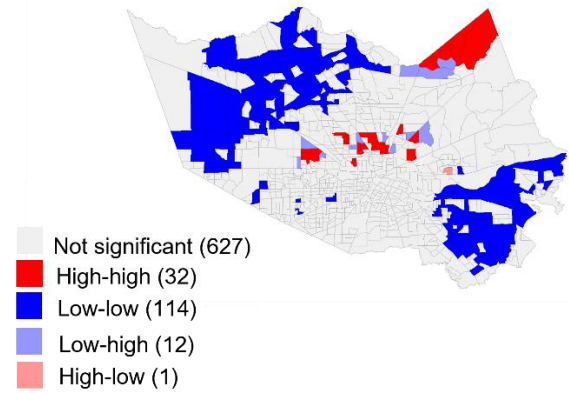
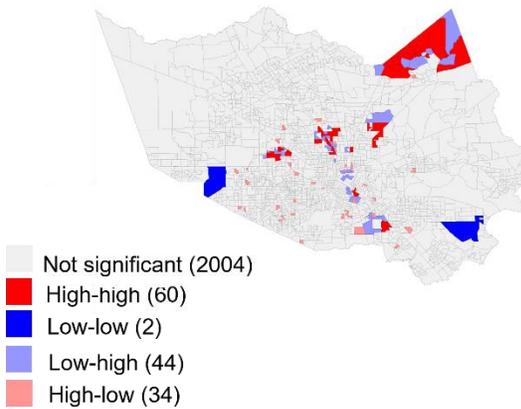
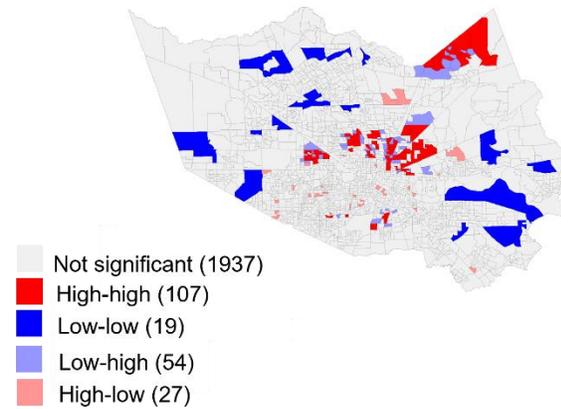

**Fig. 5. Harris County spatial maps of (a) baseline reports at the census tract level, (b) expanded reports at the census tract level, (c) baseline report at the census block group level, and (d) expanded reports at the census block group level.**

Fig. 6 compares the *Baseline Reports* (6(a)) at the census tract level and at the census block group level (6(c)) with the *Expanded Reports* at the census-tract level (6b) and the census block group level (6(d)) during Hurricane Ida. The same pattern can be seen with Hurricane Ida as with Tropical Storm Imelda. The *Expanded Reports* provide more statistically significant clusters that can offer more insights when just only looking at the intuitive Storm category.

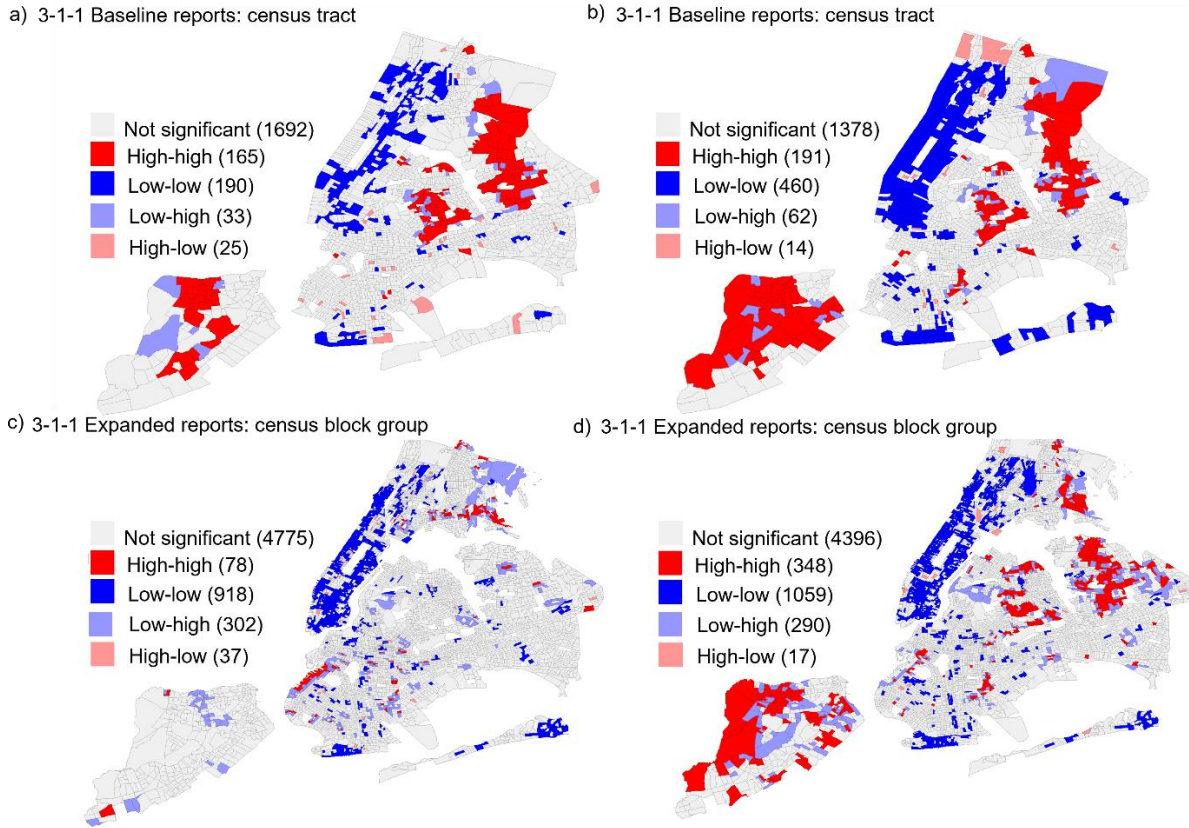

**Fig. 6. New York spatial maps of (a) baseline reports at the census tract level, (b) expanded reports at the census tract level, (c) baseline report at the census block group level, and (d) expanded reports at the census block group level.**

By expanding the topics of 3-1-1 reports, this research addresses potential sample bias and demonstrates the importance of mitigating sample bias in crowdsourced data. This research will use the expanded reports (referred to as 3-1-1 reports from this point forward). Categories at both the census tract and census block group level for further analysis in section 5.3 and 5.4.

## 5.3 Identifying Data Imbalances by Examining Spatial Biases

We examined spatial bias by performing the Global Moran's I test at the census tract and census block group level to assess the extent that geographical level of aggregation impacts the spatial results. First, the Global Moran's I was performed, and once statistically significant results are provided, the LISA test was performed on FEMA damage claims, 3-1-1 reports, and Waze reports. The LISA test identifies overlapping patterns between the three datasets to assess if damaged areas received attention through crowdsourced reporting. To further investigate spatial bias, the research then performs the BI-LISA test on FEMA damage claims with 3-1-1 reports and FEMA damage claims with Waze reports. The BI-LISA test identifies data imbalance by identifying areas of high damage with low reports and areas of low damage with high reports. To further assess generation behind the clusters of the BI-LISA test, the research performs the one-way ANOVA test to assess which social demographic trait is statistically different among the four clusters generated by the BI-LISA test.

*5.3.1 Local Cluster Assessment for Indicating Spatial Biases*

Table 4 shows the statistically significant Global Moran's I value for FEMA damage, 3-1-1 reports, and Waze reports for Hurricane Ida and Tropical Storm Imelda at the census tract level and census block group level. Moreover, the positive values for the Global Moran's I test indicate that neighboring census tracts and census blocks share a similar experience with reporting or damage patterns. The statistical significance of the study areas as a whole allows for both study areas to be decomposed into local clusters using the LISA statistic since the patterns are not random, as indicated by the Global Moran's I test.

**Table 4. Global Moran's I results for Hurricane Ida and Tropical Storm Imelda.**

|  | Census tract level | | Census block group level | |
| --- | --- | --- | --- | --- |
| Data sets | Hurricane Ida | Tropical Storm Imelda | Hurricane Ida | Tropical Storm Imelda |
| **FEMA damage** | 0.111** | 0.342*** | 0.037** | 0.447*** |
| **3-1-1 reports** | 0.355*** | 0.247*** | 0.001* | 0.229*** |
| **Waze reports** | 0.239*** | 0.604*** | 0.018* | 0.503*** |

*$p<0.05$, **$p<0.01$, *** $p<0.001$*

Fig. 7 presents the statistically significant clusters during Tropical Storm Imelda, generated by the LISA test at the census tract level for FEMA damage claims (7(a)), 3-1-1 reports (7(b)), and Waze reports (7(c)). Areas 1 and 2 in Fig. 7(a) highlight the low FEMA damage claims, 3-1-1 reports, and Waze reports. The consistency of these patterns shows that there is no evidence of data imbalance in these areas. Area 3, however, shows high damage areas and a small cluster of high–high 3-1-1 reports in an area with low–low Waze report clusters. This indicates a form of spatial bias, as certain areas are receiving attention from crowdsourced reporting from one data set, while the same area does not share similar attention with another data set, yet this area has high FEMA damage claims as seen in area 3 in Fig. 7(a). Finally, area 4 shows a large concentration of high–high Waze reports in Downtown Houston, showing that most roads were flooded due to Tropical Storm Imelda. The spatial results presented show how to mitigate data imbalance by assessing and cross validating different data sets and examining their spatial patterns to holistically look at the impact of the disaster.

At the census block group level, Fig. 8(d)–8(f) shows overlapping patterns with the spatial clusters at the census tract level. Similar to Fig. 7, a reduction in the reliability of flood mapping can be seen at a smaller geographical scale as areas 1 and 2 in Fig. 7(e) shows fewer 3-1-1 reports. Notably, the statistically significant clusters of crowdsourced data at smaller geographical scale are reduced, highlighting those smaller geographical aggregations impose a reduction of data. This reduction shows the sensitivity of flash flood mapping to the spatial level of aggregation and the fact that a smaller scale of aggregation could introduce spatial biases (blind spots) in mapping flash flood damages using crowdsourced data.

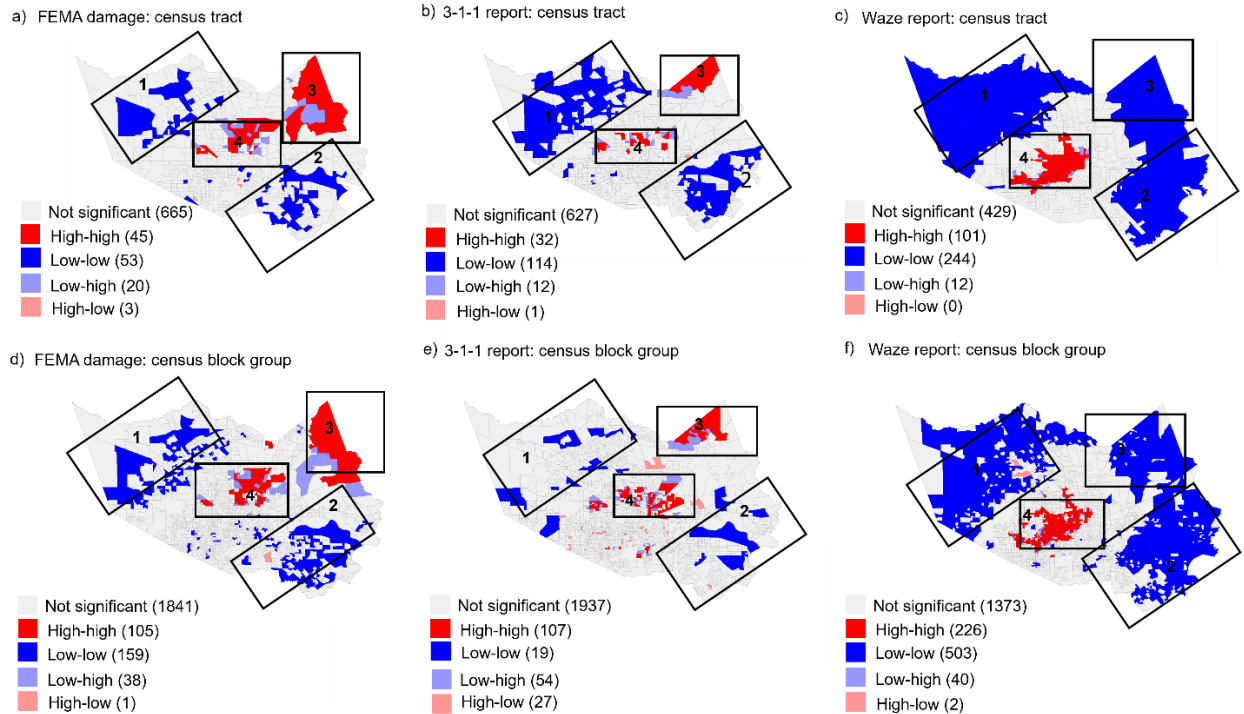

**Fig. 7.** LISA Test of (a and d) FEMA damage, (b and e) 3-1-1 reports, and (c and f) Waze reports at the census tracts (a–c) and census block group (d–f) for Harris County during Tropical Storm Imelda.

Fig, 8 shows the statistically significant local clusters for Hurricane Ida generated by the LISA test at the census tract and census block group level. As shown in Fig. 8(c), area 1 has a large cluster of high–high Waze reports implying these areas were flooded, yet they have a large concentration of low–low 3-1-1 report clusters. This highlights a form of spatial bias as there is one data set that shows low flooding (3-1-1 reports), yet another data set (Waze reports) indicates flooding, all the while having little FEMA damage claims in this area. If disaster planners only look at one source of information for flooding, they could be missing key information. Therefore, to improve situational awareness during flash floods, it is important to examine spatial biases in crowdsourced data. Area 2 had similar patterns for all three data sets, showing that this area was severely impacted by Hurricane Ida as FEMA damage, 3-1-1 reporting, and Waze reporting had an abundance of high-high Clusters. Area 3 shows similar patterns for FEMA damage and 3-1-1 reports, but there is a lack of Waze reports.

When examining the datasets at the census block group level, Fig. 8(d) shows that FEMA damage has overlapping high-high clusters in area 3 with 3-1-1 reports, but area 3 in Fig. 8(f), shows low-low clusters of Waze reports in area 3 at the census block group level in area 3. Comparing 3-1-1 reports in area 1 at the census tract (Fig. 8(b)) and census block group resolution, in area 1 in Fig. 8(b) (census tract level) and 8e (census block group level), these areas show a cluster of 3-1-1 reports; however, unlike at the census tract level in area 1, Fig. 8(f), at the census block group level, does not have a corresponding cluster of high-high Waze reports. Area 2 in Fig. 8(f) (Waze reports at the census block group level) does not have the cluster of high-high Waze reports that was noted at the census tract level (Fig. 8(c)). This result further highlights spatial bias as changing the geographical aggregation can impact the results. To mitigate spatial bias, it is important for future research to be mindful of the geographical level of aggregation that is used and gain an understanding of how the aggregation impacts the nature of the dataset. For example, a smaller geographical aggregation (census blocks) caused a loss of Waze reports' reliability for flood mapping but not 3-1-1 reports. This shows that the data sets used have their own unique characteristics, as Waze reports focuses more on roads, while 3-1-1 reports is at the household level.

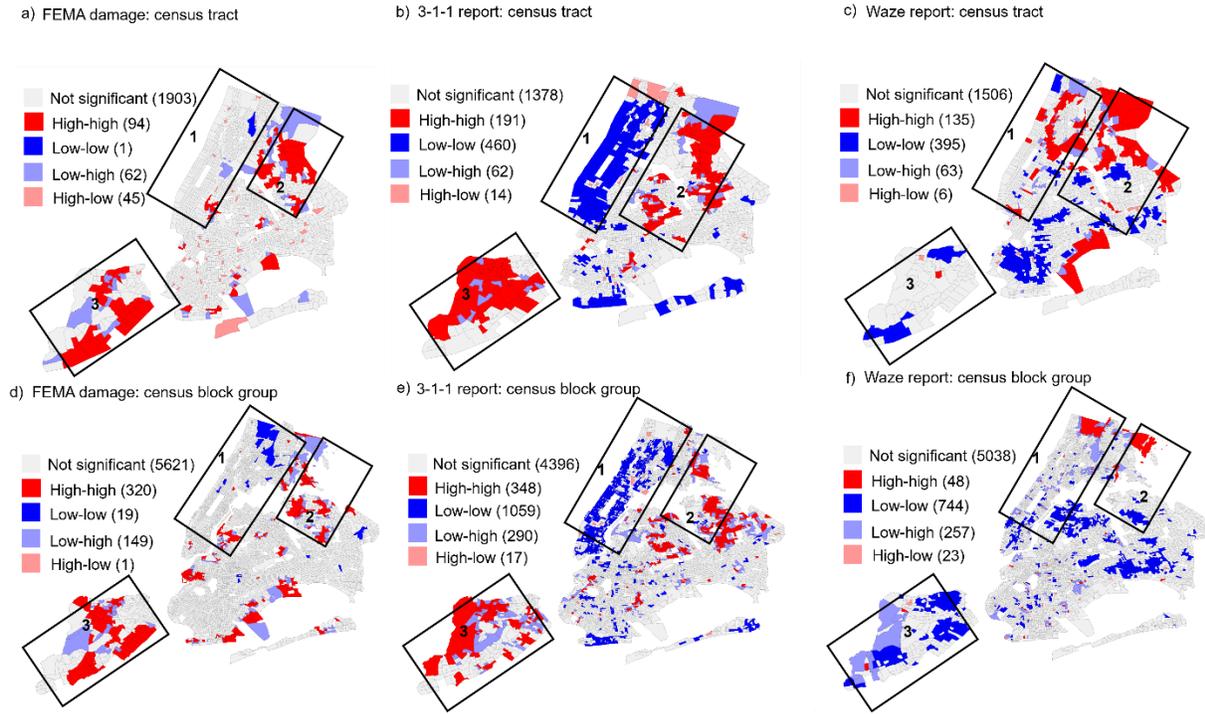

**Fig. 8. LISA Test of (a and d) FEMA damage claims, (b and e) 3-1-1 Reports, and (c and f) Waze Reports at the census tract level (a–c) and census block group level (d-f) for New York City during Hurricane Ida.**

### 5.3.2  *Examination of Data Imbalance Clusters using Multivariate Analysis*

To further investigate data imbalance, we performed the BI-LISA test. First, the Global Moran's I statistic was performed on FEMA damage claims with 3-1-1 reports and FEMA damage claims with Waze reports at both geographical aggregations during Tropical Storm Imelda and Hurricane Ida. The results are highlighted in Table 5, showing that all Moran's I values are statistically significant. Moreover, during Hurricane Ida, at the census block group level, FEMA damage with Waze reports had a negative coefficient, meaning that neighboring census block groups did not share similar experiences with damage and reporting. The remaining coefficients are positive, which indicates neighbors share a similar experience with damage and reporting extent. The statistical significance of all Moran's I values allows the research to perform the BI-LISA test.

**Table 5. Global Moran's I results for BI-LISA parameters**

|  | **Census Tract level** | | **Census Block Group level** | |
| --- | --- | --- | --- | --- |
|  | **Hurricane Ida** | **Tropical Storm Imelda** | **Hurricane Ida** | **Tropical Storm Imelda** |
| **FEMA damage with 3-1-1 reports** | 0.043*** | 0.044*** | 0.004* | 0.188*** |
| **FEMA damage with Waze reports** | 0.044*** | 0.067*** | -0.004* | 0.049*** |

*$p<0.05$, **$p<0.01$, *** $p<0.001$

Fig. 9 shows the results of the BI-LISA analysis of FEMA damage with 3-1-1 reports during Tropical Storm Imelda in Harris County at the census tract level (9(a)) and census block group level (9(c)), and Hurricane Ida in New York City at the census tract level (9(b)) and census block group level (9(d)). The BI-LISA results reveal spatial bias as it breaks down the geography into four categories: high damage-high reports, low damage-low reports, high damage-low reports, and low-damage reports.

The results show that there is data imbalance in area 1 and area 2 (Harris County) indicated by the low-high clusters in Fig. 9(a) and 9(c). This means that communities in these areas may experience damage but did not have flood insurance during Tropical Storm Imelda, therefore there was reporting of flood impacts but low FEMA damage estimation. Moreover, Fig. 9(c) shows a concentration of high-low cluster above the downtown area at the census block group level. This could indicate that the populations in this community might not have had the means to report. When examining Hurricane Ida (New York City), the results show that areas 2 and 3 at the census track and census block levels contain low-high clusters, which could indicate data imbalance. Area 1 in Figs. 9(b) and 9(d) has overlapping clusters of low-low reports which indicates a lack of data imbalance, as these communities had low damage and felt little need to report.

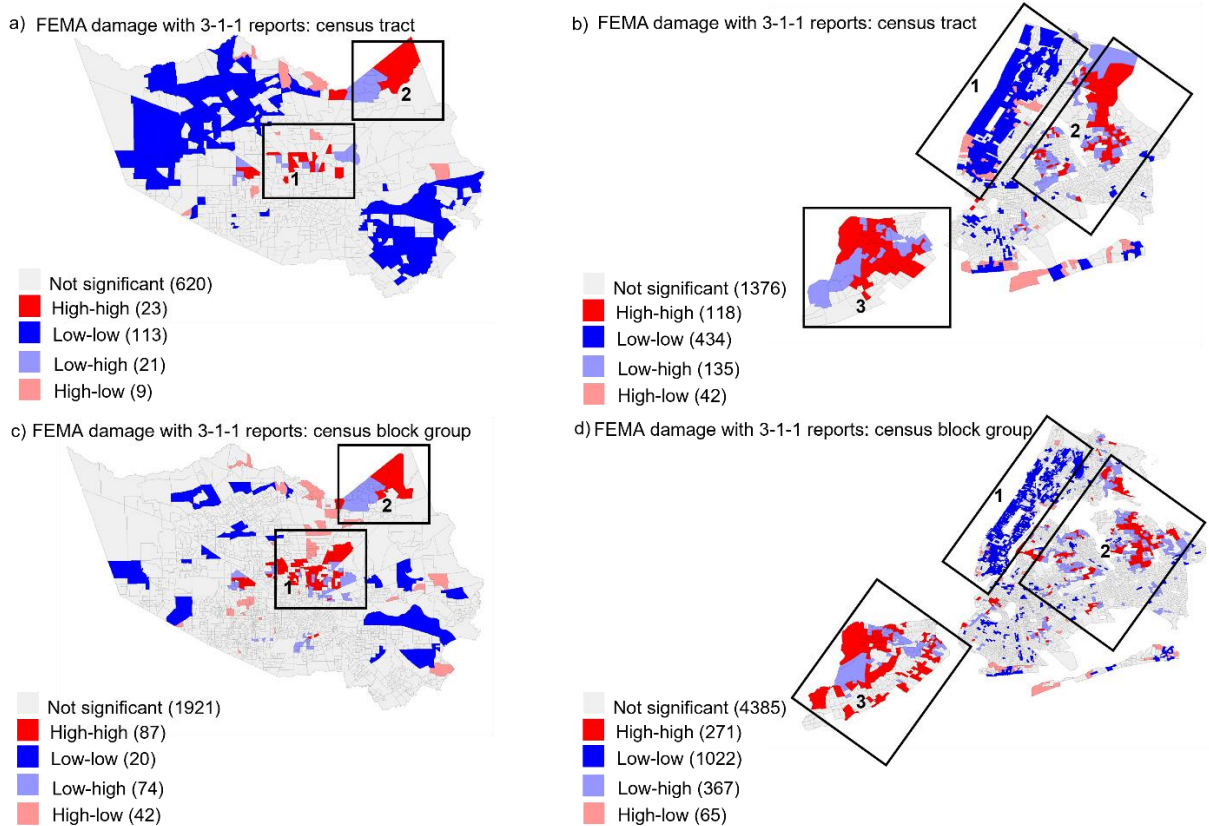

**Fig. 9. BI-LISA test for FEMA Damage with 3-1-1 reports during Tropical Storm Imelda in Harris County at the (a) census tract level, (c) census block group level. BI-LISA test for FEMA damage in New York City during Hurricane Ida at the (b) census tract level, (d) census block group level.**

Fig. 10 shows the results of the BI-LISA test for FEMA damage with Waze reports during Tropical Storm Imelda in Harris County at the census tract level (10(a)) and census block group level (10(c)) and during Hurricane Ida in New York City at the census tract level (10(b)) and census block group level (10(d)). During Tropical Storm Imelda, there was a large concentration of low-high clusters in area 4. This indicates that among the downtown area of Houston experienced data imbalance as certain communities in this area could not withstand the damage Tropical Storm Imelda imposed and were thus compelled to report. When looking at the clusters during Hurricane Ida, inconsistencies emerge between the two geographical aggregations except in area 1, where a low-high cluster is found. Moreover, in area 2 are found low-low clusters interspersed with just a few high-low clusters at the census tract level; however, these high-low clusters are lost at the census block group level highlighting spatial bias. Area 3 in Figs. (10(b)) and (10(d)) show a clear discrepancy report as a variety of clusters are statistically significant. When examining the two flooding events for data imbalance, New York has inconsistent Waze reporting patterns compared to Harris County. However, the 3-1-1 reporting patterns are consistent among both spatial groups. This highlights how spatial aggregation affects the results of different datasets and future research must be mindful when spatially aggregating data.

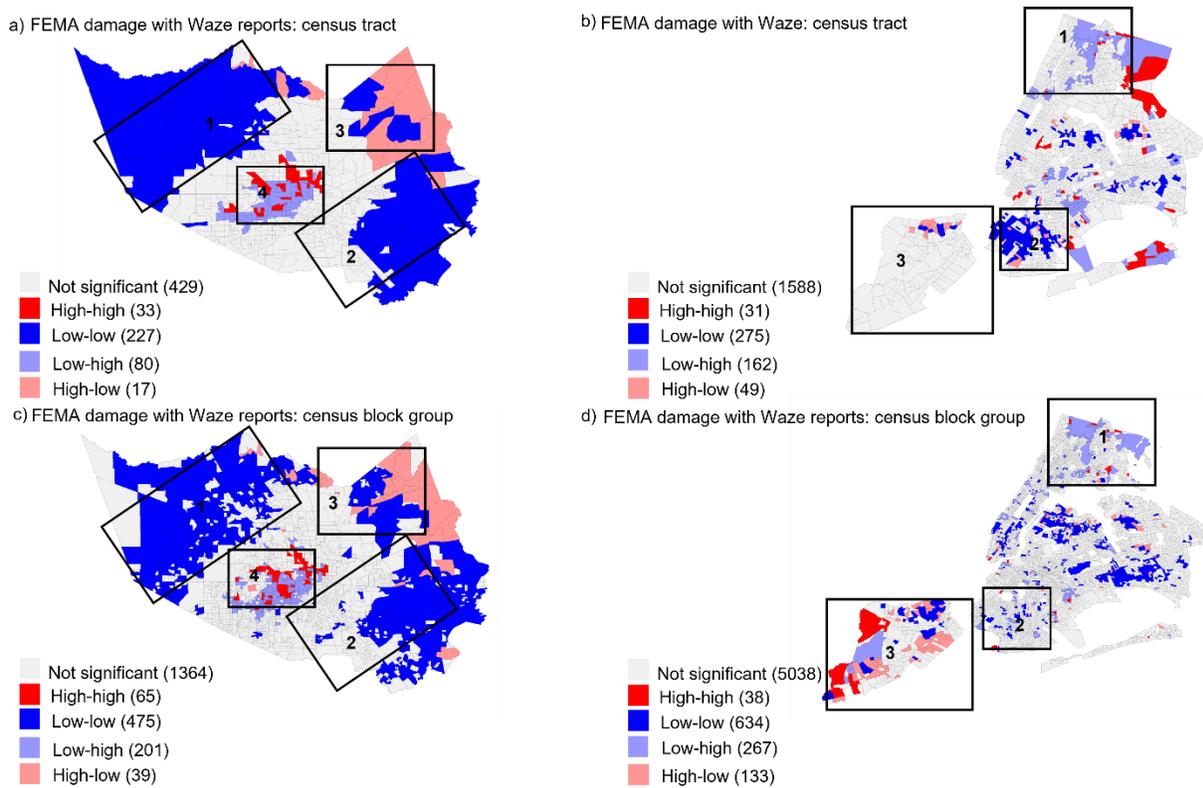

**Fig. 10. BI-LISA test for FEMA damage with Waze reports during Tropical Storm Imelda at the (a) census tract level, (c) Census Block Group level and during Hurricane Ida at the (b) census tract level, (d) census block group level.**

### 5.3.3 Assessment of Demographic Biases in Crowdsourced Data

Spatial bias has been examined using the LISA and BI-LISA statistics of different datasets. To address demographic bias and the underlying mechanisms behind the spatial clusters generated from the BI-LISA statistic, the research performs the one-way ANOVA test on all four clusters to assess how different social demographic traits are distributed among each of the four clusters. The ANOVA test was performed on the statistically significant census tracts and census block groups of the high-high, low-low, low-high, and high-low clusters from the BI-LISA test. A higher coefficient means that the social demographic traits were vary across all clusters. The social demographic traits selected are based on whether the traits have a high and significant coefficient for both flooding events.

Table 6 presents the results of the one-way ANOVA of the BI-LISA clusters at the census tract level. During Hurricane Ida, the Single parent household% share showed the largest discrepancies among clusters and the Minority% characteristic had the second largest. For Tropical Storm Imelda, Poverty% had the highest discrepancies among clusters, and Minority% had a statistically significant value. The ANOVA test demonstrates that social groups, such as Single parent households%, Poverty%, and Minority%, were susceptible to data imbalance when examining Waze reports. When examining 3-1-1 reports, Minority% and No education (NOEDU%) were susceptible to data imbalance as their ANOVA coefficient was high. For Tropical Storm Imelda, Poverty% and NOEDU% had the highest ANOVA coefficient.

Table 6. ANOVA test at the census tract level,

| Demographic characteristic | FEMA with Waze reports at the census tract level | | FEMA with 3-1-1 reports at the census tract level | |
| --- | --- | --- | --- | --- |
| | Hurricane Ida | Tropical Storm Imelda | Hurricane Ida | Tropical Storm Imelda |
| Single parent household% | 60.92*** | 2.59* | 5.822** | 2.59* |
| Poverty% | 25.56*** | 20.32*** | 1.381 | 10.75*** |
| Minority% | 43.07*** | 6.93*** | 6.277** | 4.78** |
| NOEDU% | 7.67*** | 8.13*** | 8.33*** | 11.93** |

*$p<0.05$, **$p<0.01$, *** $p<0.001$*

Table 7 presents the results of the ANOVA Test of the BI-LISA results at the census block group level. For FEMA damage with Waze Reports the only significant social group was Minority% during Hurricane Ida, while Poverty%, Minority%, and NOEDU% had large discrepancies among the BI-LISA clusters generated from Tropical Storm Imelda. When assessing FEMA damage with 3-1-1 reports, only Minority% was significant during Hurricane Ida. During Tropical Storm Imelda Single parent households%, Poverty%, and Minority% were significant.

Table 7. ANOVA Test at the census block level,

| Demographic characteristic | FEMA with Waze reports at the census block group level | | FEMA with 3-1-1 reports at the census block group level | |
| --- | --- | --- | --- | --- |
| | Hurricane Ida | Tropical Storm Imelda | Hurricane Ida | Tropical Storm Imelda |
| Single parent household% | 0.653 | 0.575 | 0.714 | 3.35* |
| Poverty% | 0.944 | 8.30*** | 0.412 | 6.609*** |
| Minority% | 17.71*** | 3.83** | 17.78*** | 6.45*** |
| NOEDU% | 0.291 | 8.98*** | 0.334 | 0.897 |

*$p<0.05$, **$p<0.01$, *** $p<0.001$*

To further illustrate data imbalance, Fig. 11 presents the box and whisker plots of Minority% at the census tract level from the results of the ANOVA test in Table 6. Minority% was selected because of the high ANOVA coefficient throughout both flooding events at both geographical aggregations. From Fig. 11(a), minorities resided in areas with high FEMA damage and high 3-1-1 reports showing that these groups faced severe impacts from Tropical Storm Imelda. Additionally, Fig. 11(b) shows that all four clusters had similar means. During Hurricane Ida, Minority populations resided in areas with low FEMA damage but high 3-1-1 and Waze reports. This highlights the possibility of minority populations not having flood insurance for their impact to be reflected in the FEMA damage data and the crowdsourced data can indicate possible significant flood impacts.

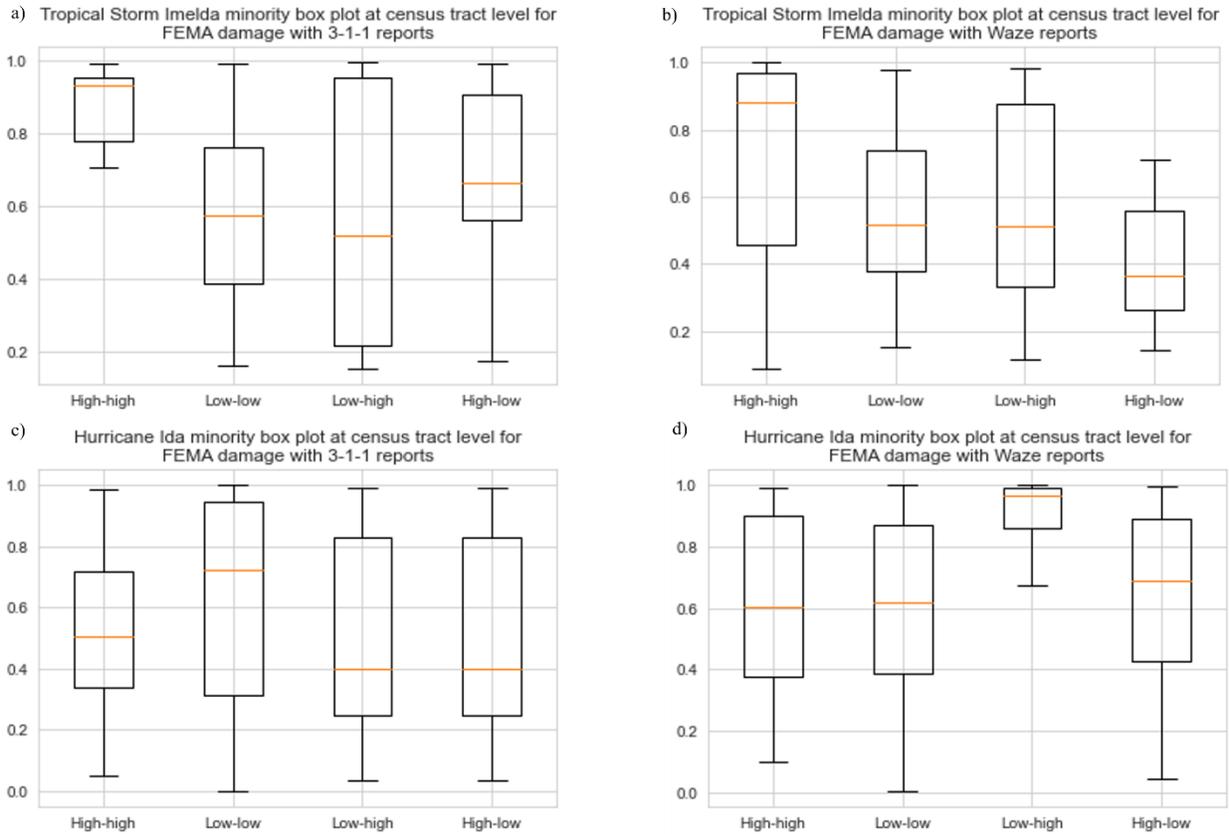

**Fig. 11. Box and whisker plot of minority populations at the census tract level for (a) Tropical Storm Imelda FEMA damage with 3-1-1 reports, (b) Tropical Storm Imelda FEMA damage and Waze reports, (c) Hurricane Ida FEMA damage with 3-1-1 reports, and (d) Hurricane Ida FEMA damage with Waze reports,**

Fig. 12 shows the ANOVA test at the census block group level. Fig. 12(a) shows that Minority populations reside in areas with low FEMA damage and low Waze reports. Fig. 9(a) and 9(c) show a reduction of low–low clusters as a smaller data aggregation were lost. Moreover, Fig. 12(b) shows a concentration of minority populations in high-high clusters followed by low FEMA damage–high Waze report clusters. In Fig. 10(a) and 10(c), it can be seen that these groups are located close to downtown Houston. During Hurricane Ida, there was a larger mean of low-high groups in Fig. 12(d).

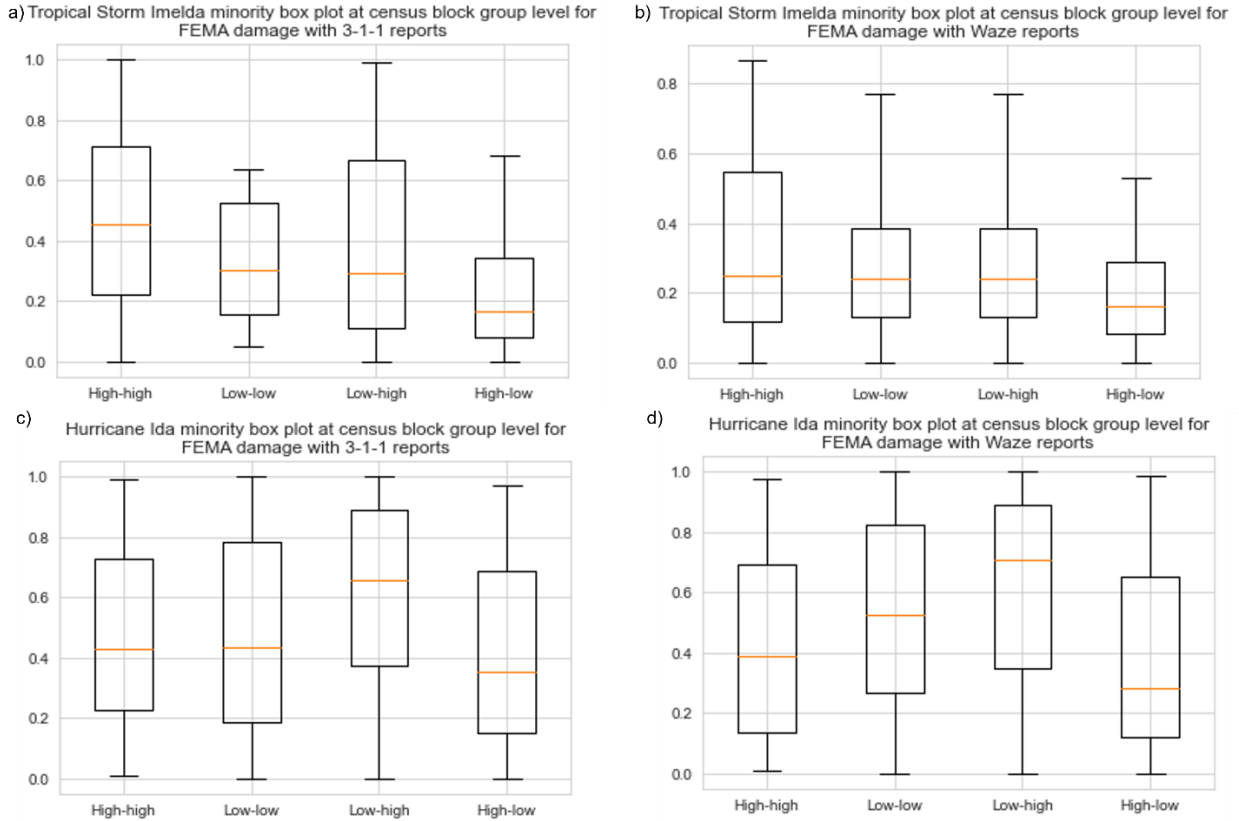

**Fig. 12. Box and whisker plot of minority populations at the census block group level for (a) Tropical Strom Imelda FEMA damage with 3-1-1 reports, (b) Tropical Storm Imelda FEMA damage and Waze reports, (c) Hurricane Ida FEMA damage with 3-1-1 reports, and (d) Hurricane Ida FEMA damage with Waze reports**

## 5.4 Identifying Data Imbalances by Examining Spatial Biases

Table 8 shows the regression results for the four models presented in Section 3 during Tropical Strom Imelda and Hurricane Ida at the census tract level and census block group level. The results for Model 1 show that population has a negative coefficient for Tropical Storm Imelda at the census tract level and a small coefficient at the census block group level. During Hurricane Ida, however, the coefficient at the census tract level was small yet statistically significant. For Model 1 at the census block group level, population loses statistical significance for both disasters. Once the damage parameter was incorporated in Model 2, the coefficient patterns stayed relatively similar at the census tract level for both disasters; however, at the census block group level for Tropical Storm Imelda, FEMA damage and population both had statistical significance, while population had a negative small coefficient. Model 3 had statistical significance for Poverty% during Tropical Storm Imelda at both geographical levels; however, at the census block group level, Poverty% had a negative coefficient implying that the lower the poverty population, the fewer reports there are. A similar pattern can be seen with Single parent households% at the census tract level, as that parameter has a negative coefficient. During Hurricane Ida, Minority% with a negative coefficient at the census tract level was revealed in Model 3; however, no social parameter was statistically significant at the census block group level. By comparing Model 3 with Models 1 and 2, it is apparent that social demographic traits have an impact on the variation of reporting rather than populous areas as both disaster events had either a negative coefficient for population or small coefficient for population. However, the social demographic traits such as Single parent household% (during Tropical Storm Imelda) and Minority% (during Hurricane Ida), had a negative coefficient, implying that these groups do not have access to reporting and suffer from population bias at the census tract level. Once all parameters are incorporated into Model 4, Single parent households% during Tropical Storm Imelda still had the negative coefficient, while Minority% loses statistical significance. This shows that single parent households are impacted by demographic bias, as this group has a negative coefficient with reports, implying that as the number of reports increase, single parent households decrease. At the census block group level, the patterns seen in previous models remained the same. During Hurricane Ida, the social parameters had no statistical significance at the census block group level. This highlights the concept that data imbalance has smaller geographical scales have little consistency.

**Table 8. Regression analysis of Waze reports at census tract and census block group level.**

| Event | Tropical Storm Imelda | | | | Hurricane Ida | | | |
|---|---|---|---|---|---|---|---|---|
| Model | Model 1 | Model 2 | Model 3 | Model 4 | Model 1 | Model 2 | Model 3 | Model 4 |
| **Census tract level** | | | | | | | | |
| Constant | 0.104*** | 0.102*** | 0.076*** | 0.098*** | 0.019*** | 0.019*** | 0.0446 | 0.054 |
| Population | -0.165*** | -0.162*** | | -0.118*** | 5.00e-6*** | 4.36e-6*** | | 4.016e-6*** |
| FEMA damage | | 0.057 | | 0.073* | | 0.387*** | | 0.374*** |
| Poverty% | | | 0.226*** | 0.187*** | | | 0.005 | 0.004 |
| NOEDU% | | | -0.041 | -0.063** | | | -0.014 | -0.049 |
| Minority% | | | -0.006 | 0.015 | | | -0.015* | 8.76e-5 |
| Single parent household% | | | -0.184*** | -0.156*** | | | 0.138*** | 0.078*** |
| $R^2$ | 0.052 | 0.054 | 0.078 | 0.106 | 0.029 | 0.077 | 0.018 | 0.084 |
| **Census block group** | | | | | | | | |
| Constant | 0.0015*** | .054*** | -0.0345 | -0.003 | 0.005*** | 0.006*** | 0.0053*** | 0.0053** |
| Population | 4.29e-07 | -3.46e-6** | | 4.3e-07 | 0.003 | 0.0029 | | -1.86e-3 |
| FEMA damage | | .122*** | | 0.176*** | | -0.010 | | -1.27e-2 |
| Poverty% | | | -0.359*** | -0.350*** | | | 1.37e-5 | 1.46e-5 |
| NOEDU% | | | 0.237*** | 0.275*** | | | -1.89e-3 | -1.35e-5 |
| Minority% | | | 0.002 | 0.005 | | | 1.55e-4 | 1.55e-4 |
| Single parent household% | | | 0.051 | 0.002 | | | -1.15e-2 | -1.05e-2 |
| $R^2$ | -0.00027 | 0.0601 | 0.0627 | 0.2977 | -0.001 | -0.002 | -0.002 | -0.0005 |

*$p<0.05$, **$p<0.01$, *** $p<0.001$

Table 9 shows the regression model of 3-1-1 reports during Tropical Strom Imelda and Hurricane Ida at the census tract and census block group level. Model 1 shows that population has a negative statistical significance for Tropical Storm Imelda at the census tract level and no statistical significance at the census block group level. During Hurricane Ida, Model 1 showed a small statistically significant coefficient at the census tract level, and a negative coefficient at the census block group level. These results show that 3-1-1 reports have occurred in less populated areas. Once the damage parameter was incorporated into Model 2, the population size coefficients stayed relatively the same for both disasters at all geographical scales. The FEMA damage claim parameter had statistical significance for both models except during Hurricane Ida at the census block group level. Model 3 examines the social parameters and at the census tract level during Tropical Strom Imelda: all parameters were statistically significant except NOEDU%. Moreover, Minority% and Single Parent Household% had negative coefficients. At the census block group level, all parameters had a positive coefficient. During Hurricane Ida, at the census tract level, only Minority%— with a negative coefficient—and Single parent household% had statistical significance. However, at the census block group level, only NOEDU% had statistical significance. By comparing Model 3 with Models 1 and 2, we find that social demographic traits impacted the variation of crowdsourced reporting. For example, population size has a negative coefficient or nearly zero coefficient in Models 1 and 2; however, demographic traits such as Minority% and Single parent households% had a negative coefficient during Tropical Storm Imelda. During Hurricane Ida, Minority% had a negative coefficient in Model 3. This shows that Single Parent Households and Minority groups suffer from demographic bias as these demographic groups are not represented on crowdsourced platforms as evidence by their negative coefficient. Model 4 includes all parameters and shows that Minority% loses statistical significance during Tropical Strom Imelda at the census tract level, but at the census block group level this parameter gained statistical significance. During Hurricane Ida, all parameters followed the patterns from the previous models at the census tract level; however, at the census block group level, none of the social parameters were statistically significant. This shows how a smaller geographical aggregation can lead to a loss of data, causing data imbalances.

Table 9. Regression analysis of 3-1-1 reports at census tract and census block group level.

| Event | Tropical Storm Imelda | | | | Hurricane Ida | | | |
|---|---|---|---|---|---|---|---|---|
| Model | Model 1 | Model 2 | Model 3 | Model 4 | Model 1 | Model 2 | Model 3 | Model 4 |
| **Census tract level** | | | | | | | | |
| Constant | 0.0220*** | 0.020*** | 0.026*** | 0.028*** | -0.001 | -0.0009 | -0.013 | -0.009 |
| Population | -0.041*** | -0.039*** | | -0.016 | 1.361e-6*** | 7.56e-7** | | 4.74e-7 |
| FEMA damage | | 0.059*** | | 0.058*** | | 0.361*** | | 0.340*** |
| Poverty% | | | 0.091*** | 0.086*** | | | -0.021 | -0.015 |
| NOHSDP% | | | 0.005 | -0.004 | | | 0.047 | 0.029 |
| Minority% | | | -0.040*** | 0.005 | | | -0.024*** | -0.014*** |
| Single parent household% | | | -0.042** | -0.039** | | | 0.102*** | 0.066*** |
| R^2 | 0.014 | 0.027 | 0.057 | 0.07 | 0.0092 | 0.194 | 0.0507 | 0.221 |
| **Census block group** | | | | | | | | |
| Constant | 0.0015*** | 0.0012*** | -0.0302* | -0.029* | 0.024*** | 0.023*** | 0.0154*** | 0.022*** |
| Population | 4.29e-07 | 3.10e-7 | | 4.3e-07 | -0.056*** | -0.0576*** | | -0.053*** |
| FEMA damage | | 0.176*** | | 0.176*** | | 0.074 | | 0.087 |
| Poverty% | | | 0.0601* | 0.067* | | | -3.812e-5 | -1.72e-5 |
| NOEDU% | | | 0.038** | 0.032* | | | 0.006* | -0.016 |
| Minority% | | | 0.009 | 0.014** | | | 0.026 | 0.027 |
| Single parent household% | | | | 0.004 | | | 0.038 | 0.011 |
| R^2 | -0.00027 | 0.0601 | 0.005 | 0.2977 | 0.0029 | 0.00308 | 0.0004 | 0.0027 |

*$p<0.05$, **$p<0.01$, *** $p<0.001$

## 6. Concluding Remarks

The objective of this study was to examine the type of biases that occur in crowdsourced reports to mitigate data imbalances for improving situational awareness. To achieve this objective, three types of biases were assessed: 1) sample bias, 2) spatial bias, and 3) demographic bias. To addresses these biases, we used spatial analysis techniques for addressing sample bias and spatial bias and a regression model to address demographic bias for the following flooding events: 1) Tropical Storm Imelda in the Harris County area in 2019, and 2) Hurricane Ida in New York in 2021. The results showed three types of data bias in crowdsourced data that often occur during emergency management, and we assessed how these biases can inhibit reliable mapping of flash flood impacts. While the number of studies leveraging crowdsourced data for disaster situational awareness is growing, limited attention has been paid to data imbalance and biases in these datasets. This study and its findings provide important insights to researchers, emergency managers, and public officials who utilize crowdsourced data for crisis situational awareness and reveal the nature of biases in these datasets. Moreover, these decision makers can implement the analytical framework presented to identify more vulnerable communities who are prone to flooding and can use the data from an unbiased perspective to create a more equitable recovery process. Hence, the study outcomes move us closer to better understanding of nature of biases in different crowdsourced data and ways to mitigate them to get the most out of these crowdsourced data for better situation awareness in crises.

**ACKNOWLEDGEMENTS**

The authors would like to acknowledge funding support from the Texas A&M X-Grant Presidential Excellence Fund, as well as the National Science Foundation CRISP 2.0 Type 2 No. 1832662. The authors would also like to acknowledge. Any opinions, findings, and conclusions or recommendations expressed in this research are those of the authors and do not necessarily reflect the views of the funding agencies.

**DATA AVAILABILITY**

The data that support the findings of this study are available from Waze, but restrictions apply to the availability of these data, which were used under license for the current study. The data can be accessed upon request submitted to the data provider. Other data (3-1-1 and census data) used in this study are all publicly available.

**DECLARATION OF COMPETING INTERESTS**

The authors declare that they have no known competing financial interests or personal relationships that could have appeared to influence the work reported in this paper.